\documentclass[aps,prl,twocolumn,superscriptaddress,floatfix]{revtex4-2}
\usepackage{amsfonts}
\usepackage{amssymb,amstext,amsmath}
\usepackage{graphicx}
\usepackage{bm}
\usepackage{color}
\usepackage{latexsym}
\usepackage{mathtools}
\usepackage[colorlinks=true, allcolors=blue]{hyperref}
\usepackage{braket}

\newcommand{\vparallel}{\hspace{0.1cm}\rule[-0.2mm]{0.8pt}{2.4mm} \hspace{0.15cm} \rule[-0.2mm]{0.8pt}{2.4mm} \hspace{0.1cm}}
\newcommand{\hparallel}{\hspace{0.1cm}\rotatebox{90}{\rule[-0.2mm]{0.8pt}{2.2mm} \hspace{-0.1cm} \rule[-0.2mm]{0.8pt}{2.2mm}} \hspace{0.1cm}}

\usepackage{multibib} 
\usepackage{setspace} 

\usepackage{titlesec}
    \titleformat*{\section}{\large
    \bfseries}
    \titleformat*{\subsection}{\bfseries}
    \titleformat*{\subsubsection}{\itshape}

\begin{document}
\title{Shaping Magnetic Order by Local Frustration for Itinerant Fermions on a Graph}
\author{Revathy B. S}
\author{Shovan Dutta}
\affiliation{Raman Research Institute, Bangalore 560080, India}


\begin{abstract}
Kinetic magnetism is an iconic and rare example of collective quantum order that emerges from the interference of paths taken by a hole in a sea of strongly interacting fermions. Here the lattice topology plays a fundamental role, with odd loops frustrating ferromagnetism, as seen in recent experiments. However, the resulting magnetic order on a general graph has remained elusive. Here we systematically establish a general principle: that local frustration centers bind singlets while sharing a delocalized hole. This collective effect|absent in exchange magnetism|extends from rectangular grids to random graphs, producing sharp and predictable variation with tunable frustration measures. Our findings demonstrate that one can shape the spin order and tune the net magnetization by embedding kinetic frustration, opening ways of spatially resolved quantum control of many-body systems. We outline a protocol to realize some of the key findings in existing cold-atom setups.
\end{abstract}

\maketitle

Frustrated magnetism is a classic setting in which the lattice topology dictates collective ordering, giving rise to novel phases of matter such as spin liquids and spin ice \cite{intro_FM_book, diep_book, balents2010spin}. In the more commonly studied case, frustration arises from competing spin-exchange interactions defined on the bonds of a lattice, which cannot all be simultaneously satisfied \cite{moessner2006geometrical}. A distinct type of frustration arises from destructive quantum interference of paths taken by mobile spins, which increases the kinetic energy for certain spin configurations \cite{haerter_shastry_triangular_2005, mattis_book}. Unlike exchange coupling, this kinetic magnetism has no classical counterpart. How such a purely quantum-mechanical order is governed by the loop structure of the lattice is a fundamental question that has remained elusive for several decades.

Both exchange and kinetic magnetism are found in the Hubbard model \cite{keeling_notes, tasaki_book, hubbard_review} which, in the limit of strong on-site repulsion, describes hopping of spin-$\uparrow$ and spin-$\downarrow$ fermions along with an antiferromagnetic spin exchange. The hopping is suppressed when each site has one spin (half filling), but dominates in the presence of a hole. Every hop by the hole rearranges the spins and changes the sign of the wavefunction. Thus, moving the hole around a loop with even number of sides gives a constructive interference if the spins are aligned. This effect makes the ground state ferromagnetic on generic bipartite lattices, as was shown first by Nagaoka \cite{nagaoka1966ferromagnetism}. In contrast, an odd loop frustrates this order \cite{haerter_shastry_triangular_2005}, favoring a singlet (spin zero) ground state for one triangle \cite{triangular_cactus_2023}. For extended regular lattices with frustration, the ordering varies with geometry|while a triangular lattice gives rise to a $120^{\circ}$ classical antiferromagnet \cite{haerter_shastry_triangular_2005, numerical_manuel_2014}, special corner-sharing geometries such as a triangular cactus \cite{triangular_cactus_2023} or a pyrochlore lattice \cite{glittum2025resonant, poilblanc2004} can stabilize a spin liquid of singlets. Recent experiments have supported these findings on a square plaquette \cite{experiment_quantum_dots_2020} and a triangular lattice \cite{experiment_greiner_2024, experiment_bakr_2024}. However, the path interferences are considerably more complex on a general graph, where both even and odd loops exist side by side. While one expects the ferromagnetic state to be destabilized \cite{numerical_hanisch_1995, numerical_manuel_2014, numerical_gazza_2017}, there is no understanding of the resulting magnetic order, despite the possibility of realizing such connectivities in optical traps \cite{Gross2021, nixon2024individually, GonzlezCuadra2023} or with qubit encodings~\cite{fermionsimul_bravyi_kitaev_2002, fermionsimul_verstraete_cirac_2005, fermionsimul_steudtner_wehner_2019}.

This is also a fundamental question from the point of view of network science \cite{strogatz_review_2001, newman_book}, where a principal finding of a growing number of studies is that network structure can radically alter collective phenomena such as disease spreading, percolation, and synchronization \cite{dynamics_on_networks_book, critical_phenomena_review_2008, explosive_phenomena_2019}. However, these findings are limited to classical systems or, in a few cases, quantum systems with a classical analog \cite{quantum_networks_2024}. In particular, studies of frustration have focused on social balance \cite{social_balance_2014} and Ising spin glass physics \cite{critical_phenomena_review_2008}.

Here, we explore the impact of local frustration on kinetic magnetism using a bottom-up approach. By adding frustration centers (diagonal bonds) to a rectangular grid one at a time, we establish that they locally bind singlets while preserving delocalization of the hole and the spin-polarized background, which allows one to shape the spin ordering as well as tune the net magnetization in steps of $1$ by embedding frustration (Fig.~\ref{Fig4}). Moreover, this strong, collective effect extends to random graphs, where the magnetization is anticorrelated with different measures of local frustration and exhibits a characteristic oscillation as a function of the smallest loop size (Fig.~\ref{Fig5}). Such prominent variation is in stark contrast to exchange magnetism at half filling, where both the total spin \cite{preethi2024frustrated} and the local spin alignment show a lack of sensitivity to frustration; see Supplemental Material (SM) \cite{SuppMat}. 

{\it Setup}|We consider the Hubbard Hamiltonian \cite{hubbard_review, essler2005one}
\begin{equation}
\hat{H} = -  \sum_{\langle i,j \rangle, \sigma} t_{i, j } (\hat{c}_{i\sigma}^{\dagger} \hat{c}_{j\sigma} + \hat{c}_{j\sigma}^{\dagger} \hat{c}_{i\sigma} ) + U \sum_i \hat{n}_{i \uparrow} \hat{n}_{i \downarrow},
\end{equation}
where the first term describes hopping of a fermion with spin $\sigma \in \lbrace \uparrow, \downarrow \rbrace $ between neighboring sites $i$ and $j$ with amplitude $t_{i,j} >0$, and the second term describes on-site interaction between fermions of opposite spin, with occupations $\hat{n}_{i \sigma} \coloneqq \hat{c}_{i\sigma}^{\dagger} \hat{c}_{i\sigma}$ (For $t_{i,j} < 0$, ferromagnetism is not frustrated \cite{tasaki1989extension, 15puzzle_2018, powell_balance_2017}). In order to focus on the role of network structure on kinetic magnetism, we study the limit of one hole and $U\rightarrow \infty$. The latter suppresses double occupancies so the physics is governed solely by the kinetic energy of the hole.
We define the on-site spin operators $\hat{S}^z_i \coloneqq \frac{1}{2}(\hat{n}_{i\uparrow} - \hat{n}_{i\downarrow})$,  $\hat{S}^+_i \coloneqq  \hat{c}_{i\uparrow}^{\dagger}\hat{c}_{i \downarrow}$, $\hat{S}^-_i \coloneqq \hat{c}_{i\downarrow}^{\dagger} \hat{c}_{i \uparrow}$, and collective spin operators $\hat{S}^z = \sum_i \hat{S}^z_i$, $\hat{S}^{\pm} = \sum_i \hat{S}^{\pm}_i$. The net magnetization, measured by $\hat{S}^2 = (\hat{S}^z)^2 + \frac{1}{2} (\hat{S}^+ \hat{S}^- + \hat{S}^-\hat{S}^+) $, has eigenvalues of the form $S_{\rm total}(S_{\rm total}+1)$ with $S_{\rm total} \in \{N/2, N/2-1, \dots, N/2 - \lfloor N/2 \rfloor\}$, where $N$ is the total number of fermions. The Hubbard Hamiltonian has a global rotational symmetry \cite{tasaki_book, hubbard_review} such that each energy eigenstate has a definite $S_{\rm total}$ and a degeneracy of $2S_{\rm total}+1$, which corresponds to different numbers of $\downarrow$ spins, $N_{\downarrow} \in [N/2 - S_{\rm total}, N/2 + S_{\rm total}]$. 
It is thus sufficient to compare the lowest energies for $N_{\downarrow} = N/2 - S_{\rm total}$ $\in [0, N/2]$ to find the ground-state magnetic order.

\begin{figure}[ht]
\includegraphics[width=\columnwidth]{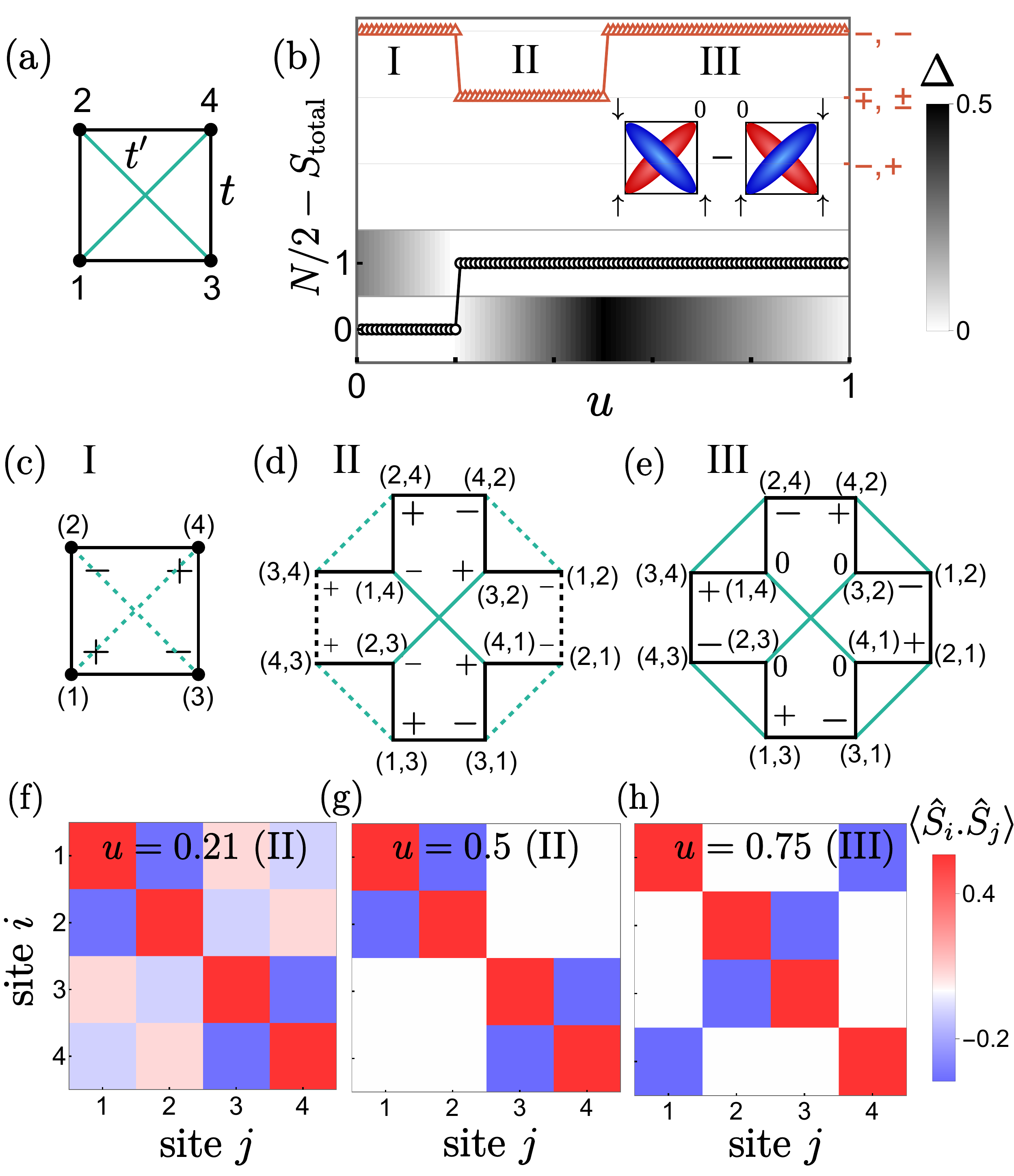}
\caption{(a) Square with tunneling $t= 1-u$ along the sides (black) and $t^{\prime}=u$ on the diagonals (green), containing $N=3$ fermions with total spin $S_{\text{total}} \in \{1/2, 3/2\}$. (b) Magnetization (circles) and reflection parity (triangles) of the ground state as a function of $u$. The spin gap $\Delta$ (minimum energy gap from the ground-state sector) is plotted in the background. The inset shows a ``cross-dimer'' structure of the ground state in region III, where blue and red ellipses denote a singlet and a delocalized hole, respectively. (c)-(e) Hilbert-space graph representation of the ground state. The vertices correspond to the position of the hole in I and those of the hole and the $\downarrow$ spin (magnon) in II and III. The signs indicate the relative amplitudes and dashed lines show frustrated edges. (f)-(h) Spin correlations showing anti-aligned spins on the vertical bonds in II and on the diagonal bonds in III.}
\label{Fig1}
\end{figure}

{\it Square with diagonal bonds}|We start by analyzing a single square with hopping $t$ along the sides and $t^{\prime}$ across the diagonals [Fig.~\ref{Fig1}(a)]. The square has reflection symmetry about the $x$ and $y$ axes leading to four parity sectors $\lbrace (+,+), (+,-), (-, +), (-,-)\rbrace$, of which $(+,-)$ and $(-,+)$ have the same energies. For $t^{\prime} = 0$, the ground state is a Nagaoka ferromagnet with $S_{\rm total} = \frac{3}{2}$ \cite{experiment_quantum_dots_2020}.

To study the effect of frustration, we set $t^{\prime} = 1-t =  u$ and plot the minimum energy sector as a function of $u$ in Fig.~\ref{Fig1}(b). For $N_{\downarrow} = 0$, the ground state is just that of a hole hopping with amplitudes $-t$ and $-t^{\prime}$, given by $\ket{\psi} = \frac{1}{2} (\hat{c}_{1\uparrow}-\hat{c}_{2\uparrow}-\hat{c}_{3\uparrow}+\hat{c}_{4\uparrow})\ket{\uparrow \uparrow \uparrow \uparrow}$ for $t^{\prime} < t$. This state is represented in Fig.~\ref{Fig1}(c) where the dotted bonds are frustrated and the alternating signs indicate the $\pi$ quasimomentum. The state has energy $E = -2t + t^{\prime}$ and remains the ground state for $u < \frac{1}{5}$, beyond which the square binds a singlet, requiring $\smash{N_{\downarrow}} = 1$. Here the system can take 12 different configurations $(i,j) \equiv \hat{S}_j^{-} \hat{c}_{i\uparrow} \ket{\uparrow \uparrow \uparrow \uparrow} $ where $i$ denotes the position of the hole and $j$ denotes that of the $\downarrow$ spin (magnon). The problem then becomes that of a single particle hopping on a 12-site Hilbert-space graph shown in Figs.~\ref{Fig1}(d) and \ref{Fig1}(e). For $t^{\prime} > t$, the frustration is minimized by the configuration in Fig.~\ref{Fig1}(e), which describes the state 
\begin{equation*}
    \ket{\times}_{\uparrow} \!=\! \big[(\hat{S}_1^- \!- \hat{S}_4^-) (\hat{c}_{2 \uparrow} - \hat{c}_{3 \uparrow}) - (\hat{S}_2^- \!- \hat{S}_3^-) (\hat{c}_{1 \uparrow} - \hat{c}_{4 \uparrow})\big] \ket{\uparrow \uparrow \uparrow \uparrow} \!,
\end{equation*}
corresponding to a singlet on one diagonal and an $\uparrow$ spin hopping on the other diagonal [inset of Fig.~\ref{Fig1}(b)], leading to diagonal singlet correlations [Fig.~\ref{Fig1}(h)]. This ``cross-dimer'' state with $(-,-)$ parity has energy  $E = -t - t^{\prime}$ and is the ground state for $u>\frac{1}{2}$. For $ \frac{1}{5} < u < \frac{1}{2}$, the ground state varies with $u$, with $E = -\sqrt{3t^2 + t^{\prime 2}} / (t + t^{\prime})$ and $(\pm,\mp)$ parity. The one with $(-,+)$ parity is sketched in Fig.~\ref{Fig1}(d) and has the form $\ket{\psi} = \ket{\vparallel} + x \ket{\hparallel}$, where 
\begin{align*}
    \ket{\vparallel} \!=\! \big[(\hat{S}_1^- \!\!-\! \hat{S}_2^-) (\hat{c}_{3 \uparrow} \!-\! \hat{c}_{4 \uparrow}) \!-\! (\hat{S}_3^- \!\!-\! \hat{S}_4^-) (\hat{c}_{1 \uparrow} \!-\! \hat{c}_{2 \uparrow})\big] \ket{\uparrow \uparrow \uparrow \uparrow} \!, \\[0.2em]
    \ket{\hparallel} \!=\! \big[(\hat{S}_1^- \!\!-\! \hat{S}_3^-) (\hat{c}_{2 \uparrow} \!+\! \hat{c}_{4 \uparrow}) \!+\! (\hat{S}_2^- \!\!-\! \hat{S}_4^-) (\hat{c}_{1 \uparrow} \!+\! \hat{c}_{3 \uparrow})\big] \ket{\uparrow \uparrow \uparrow \uparrow} \!,
\end{align*}
and $x$ increases from $0$ to $\frac{1}{5}$ as $u$ decreases from $\frac{1}{2}$ to $\frac{1}{5}$. The $\ket{\vparallel}$ component gives rise to singlet correlations on the vertical bonds, as shown in Figs.~\ref{Fig1}(f) and \ref{Fig1}(g).
 
\begin{figure}
\includegraphics[width=\columnwidth]{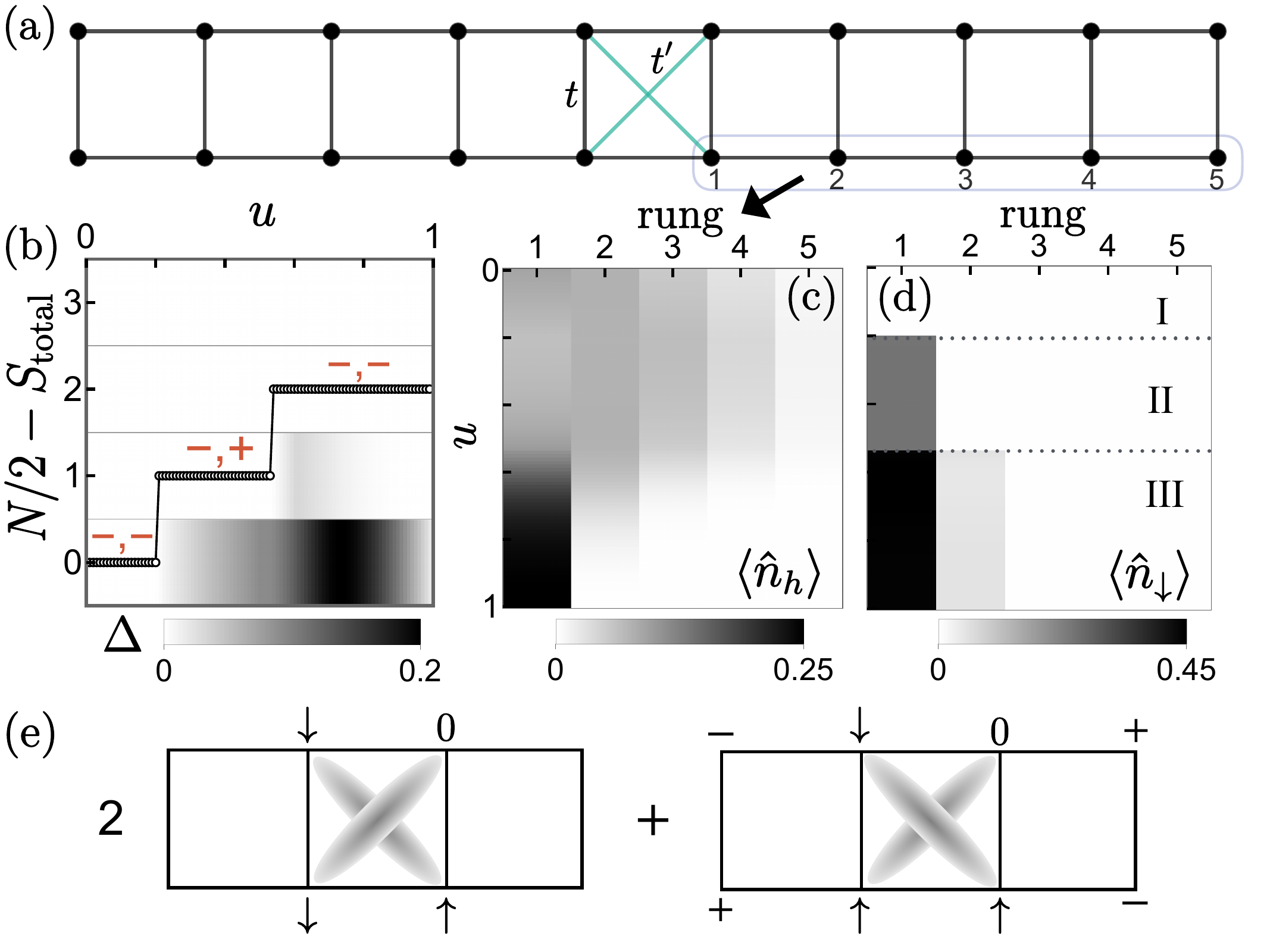}
\caption{(a) Ladder with a central frustrated plaquette. (b) Minimum energy sector and ground-state parity as a function of $u = t^{\prime} = 1-t$, with the excitation gap $\Delta$ plotted in the background. (c) Hole density $\langle \hat{n}_h \rangle$ and (d) $\downarrow$-spin density $\langle \hat{n}_{\downarrow}\rangle$, showing that the magnons are strongly bound to the center even when the hole is not. (e) Sketch of the cross-dimer type ground state for large $u$ (see text). The results are for a ladder of length $l=10$; however, they do not change beyond $l \gtrsim 4$. }
\label{Fig2}
\end{figure}

{\it Ladder with a frustrated plaquette}|Next, we embed a frustrated square at the center of a ladder [Fig.~\ref{Fig2}(a)]. From exact diagonalization, there are again three regions, except that $S_{\rm total}$ is reduced by $2$ in region III [Fig.~\ref{Fig2}(b)]. Surprisingly, the magnons are always strongly bound to the frustrated plaquette [Fig.~\ref{Fig2}(d)], even when the hole is delocalized for $u \lesssim \frac{1}{2}$ [Fig.~\ref{Fig2}(c)]. This is consistent with an effective attraction between the hole and magnons in the presence of frustration \cite{demler_attraction_2024}. The ordering is similar to that of the single square, with singlet correlations on the vertical rungs in region II and on the diagonals in region III. Due to the strong binding, the physics is unaltered by increasing the length $l$ of the ladder.

The small spread in the magnon density outside the central plaquette in region III [Fig.~\ref{Fig2}(d)] can be understood from perturbation theory in $t/t^{\prime}$. Here the hole is bound to the central plaquette, and to first order the perturbation selects two types of cross-dimer states with $(-,-)$ parity: (i) $\ket{\times}_{\downarrow}$ where the hole hops with a $\downarrow$ spin across a diagonal, and (ii) $\ket{\times}_{\uparrow,d}$ where the hole hops with an $\uparrow$ spin across a diagonal and the second $\downarrow$ spin is at a distance $d \in \{1,2,3,\dots\}$ from the central plaquette. In order to lift this degeneracy, the hole has to make three hops along a neighboring plaquette, which couples $\ket{\times}_{\downarrow}$ and $\ket{\times}_{\uparrow,1}$, selecting the ground state $2\ket{\times}_{\downarrow} + \ket{\times}_{\uparrow,1}$, as sketched in Fig.~\ref{Fig2}(e). Hence, the probability of finding a magnon outside the central plaquette is $P_{\rm out} =  \frac{1}{5}$. Note this perturbative estimate remains valid throughout region III [Fig.~\ref{Fig2}(d)]. We also show in Appendix A that the physics is similar for a square grid with $x$-$y$ symmetry and a frustrated central plaquette. Here the magnons for large $t^{\prime}$ are even more strongly bound with $ P_{\rm out} = \frac{1}{9}$.

In contrast, for Heisenberg antiferromagnetism at half filling, $S_{\rm total} = 0$ for all $u$ and the spin alignment across the diagonal bonds varies smoothly with $u$ (see SM \cite{SuppMat}).

\begin{figure}
\includegraphics[width=\columnwidth]{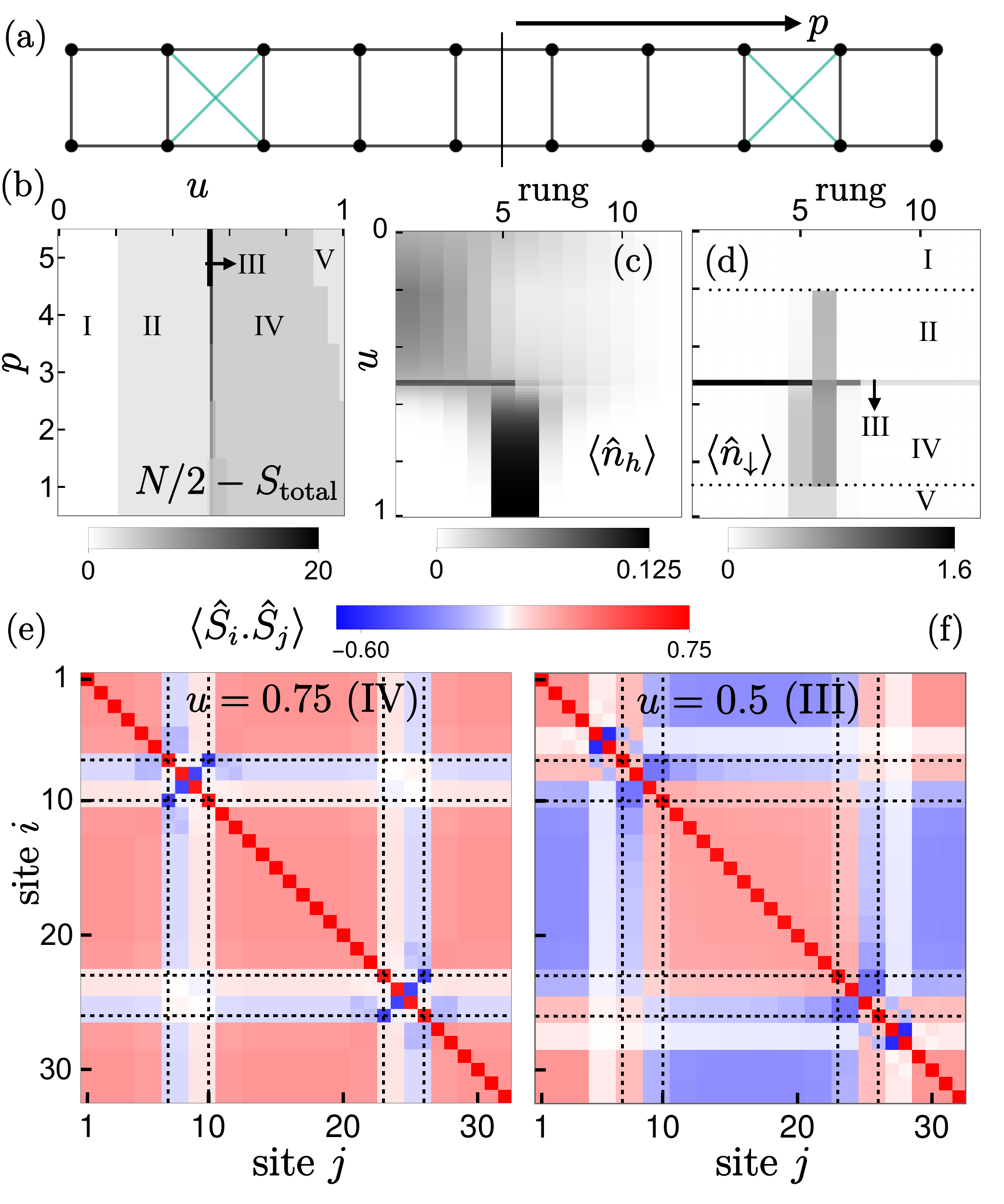}
\caption{(a) Ladder with two frustrated plaquettes at a distance $p$ from the center. (b) Phase diagram for $l=24$, where $N/2 - S_{\rm total} = 2$ (II and V), $4$ (IV), $\sim \text{min}(4p, 2l-4p)$ (III). (c) Hole density and (d) magnon density for $l = 24, p =5$. (e)-(f) Spin correlations for $l=16, p=4$, showing singlet on each diagonal in region IV and two domain walls in region~III. Sites are numbered from left to right, with odd integers for the lower leg and even integers for the upper leg, as in Fig.~\ref{Fig1}(a). The frustrated plaquettes behave collectively in III and IV.}
\label{Fig3}
\end{figure}

\begin{figure}[b]
    \includegraphics[width=\columnwidth]{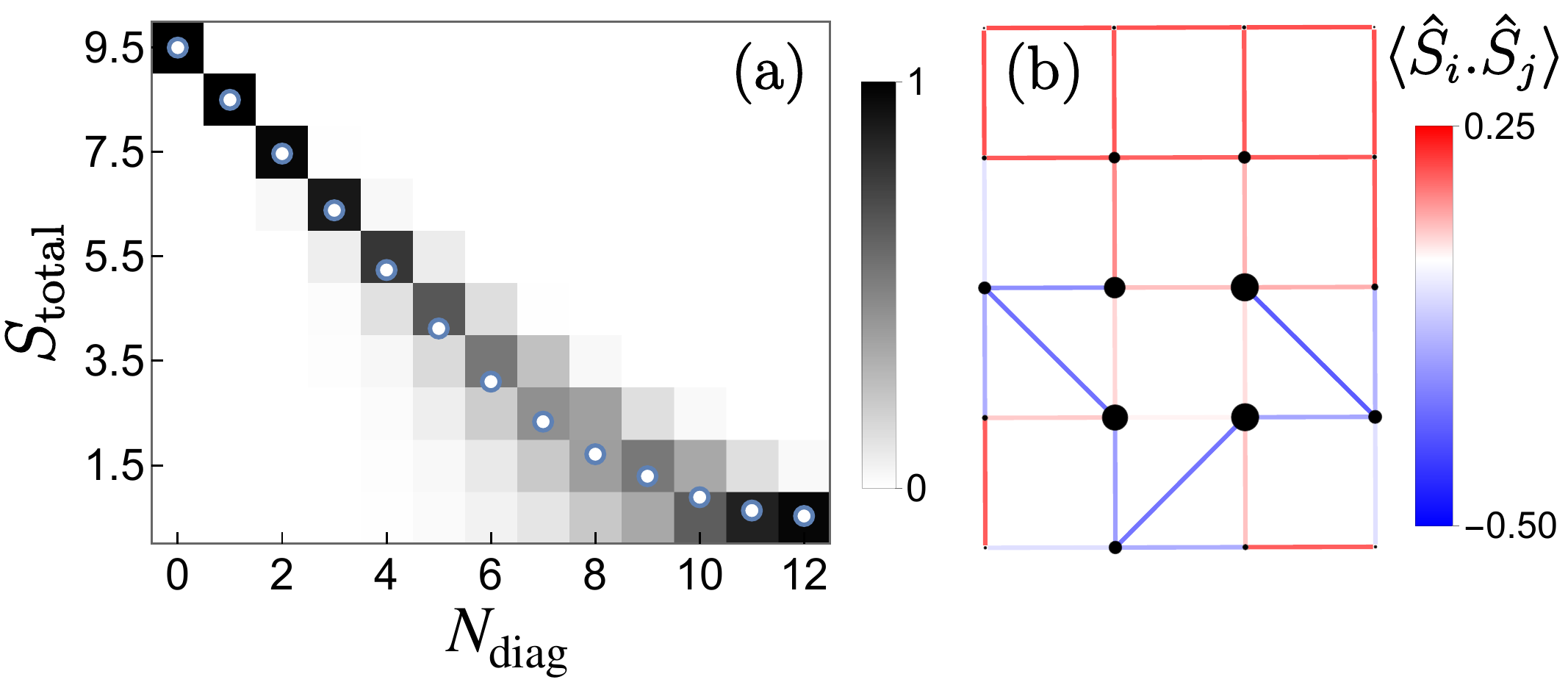}
    \caption{(a) Reduction of total spin with the number of diagonal bonds in a $4 \times 5$ grid, where the bonds are added randomly with at most one per plaquette. Dots show the average values and gray scale shows the distributions for up to $1000$ graphs. Allowing two diagonals per plaquette produces a similar variation. (b) Distribution of the hole, given by the radii of the black dots, and spin correlation on the bonds, given by the color scale, for a sample graph, showing singlets on the diagonals while the other sites are aligned. Generally, correlations between non-neighboring sites (not shown) are weaker.} 
    \label{Fig4}
\end{figure}

\begin{figure}
    \includegraphics[width=\columnwidth]{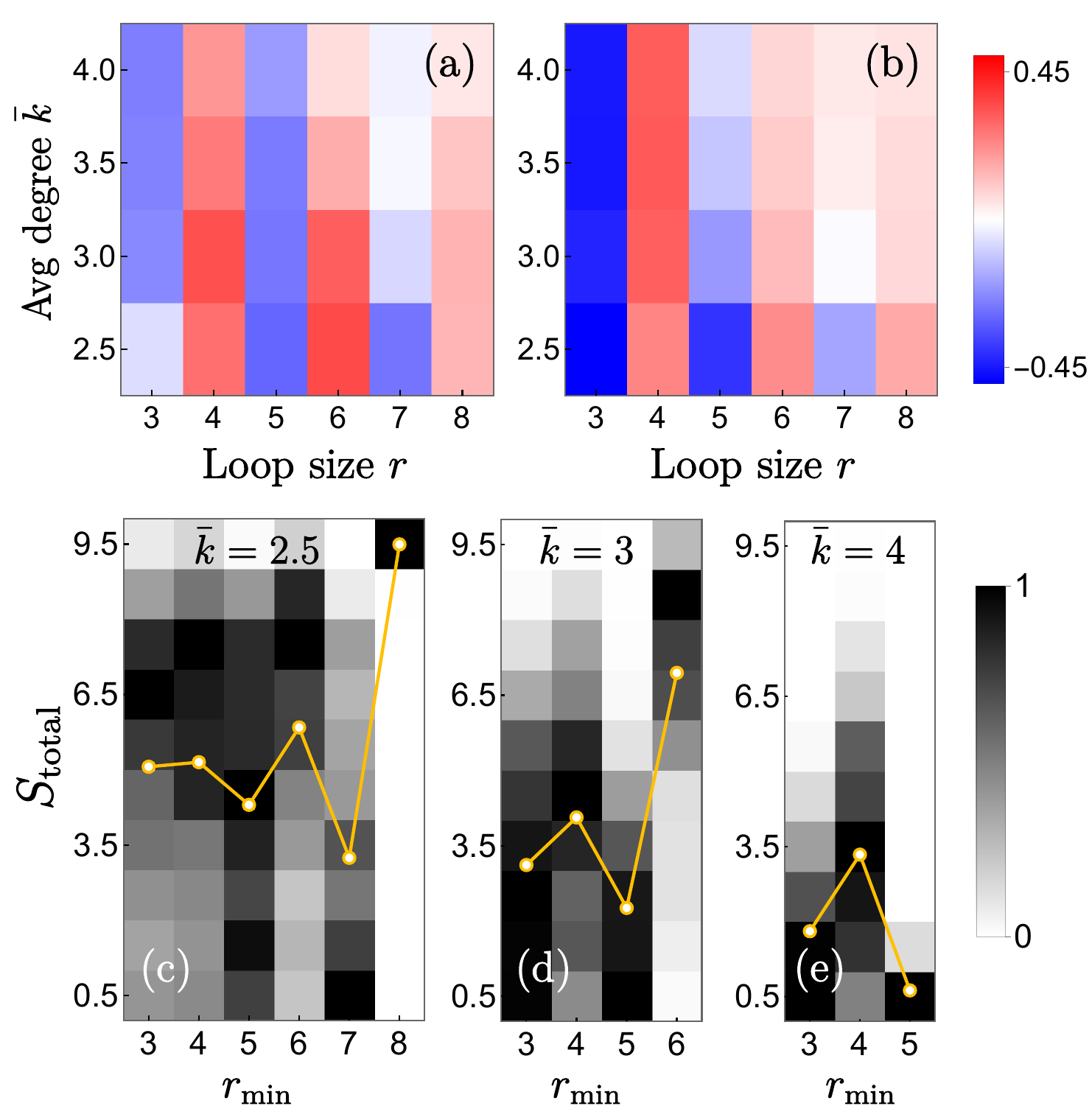}
    \caption{(a) Correlation between $S_{\rm total}$ and the number of loops of a given size $r$ in $20$-site random graphs, where $\bar{k}$ is the average number of neighbors per site. (b) Average correlation between the spin alignment across individual bonds, $\langle \hat{S}_i.\hat{S}_j \rangle$, and the number of loops that the bond is a part of, showing that the alignment is strongly governed by local frustration. (c)-(e) Oscillation of the distribution (gray scale) and average value (yellow dots) of $S_{\rm total}$ with the minimum loop size, $r_{\rm min}$, in a graph. The statistics for each parameter set are obtained from an ensemble of $1000$ nonseparable graphs.}
    \label{Fig5}
\end{figure}

{\it Ladder with two frustrated plaquettes}|We next model a ladder with two frustrated squares embedded symmetrically at a distance $p$ from the center [Fig.~\ref{Fig3}(a)]. When $p$ is sufficiently small, the hole mediates interactions between the two, producing strong collective effects.

The phase diagram for large $l$ shows $5$ different regions (Fig.~\ref{Fig3}). Two of these can be understood from the physics of the single frustrated plaquette: In region II, the hole is delocalized and each frustrated plaquette binds one $\downarrow$ spin, with singlet correlations along the vertical rungs. This is similar to region II in Fig.~\ref{Fig2}(d) and Fig.~\ref{Fig1} except that the hole and magnon densities are asymmetric across each plaquette. In region V, on the other hand, the hole is trapped in either one of the two plaquettes in a cross-dimer state, similar to region III in Fig.~\ref{Fig2}, while the other one is spin polarized (see Fig.~\ref{Fig7} for the spin correlations). In both cases, $S_{\rm total}$ is reduced by $2$.

In contrast, regions III and IV represent collective behavior. In region IV, the hole is shared by the two plaquettes, each of which binds two magnons, reducing $S_{\rm total}$ by 4. This state exhibits singlet correlations on the diagonals [Fig.~\ref{Fig3}(e)], which originates from a cross-dimer on one of the plaquettes and distorted singlets on the other, captured by the variational wavefunction
\begin{equation}
    |\psi (\alpha)\rangle = \ket{\bm{\textcolor{blue}{\times}} (\alpha)} \otimes \ket{\times}_{\downarrow} + \ket{\times}_{\downarrow} \otimes \ket{\bm{\textcolor{blue}{\times}} (\alpha)}.
    \label{eq:var_wf}
\end{equation}
Here, $\ket{\bm{\textcolor{blue}{\times}} (\alpha)}$ describes distorted singlets of the form 
\begin{equation}
    \ket{\bm{\textcolor{blue}{\times}} (\alpha)} = (S_2^- - \alpha S_3^-) (S_1^- - \alpha S_4^-) \ket{\uparrow \uparrow \uparrow \uparrow}
    \label{eq:distorted_singlets}
\end{equation}
across the diagonals of a square, where the site numbers refer to those in Fig.~\ref{Fig1}(a) and $\alpha \sim 0.6$ is the deformation parameter (see Appendix B for details). Crucially, this ordering would not be possible if the plaquettes behaved independently, as there is no reason to bind the distorted singlets in the absence of the hole. Indeed, the state is destabilized by increasing the separation $p$ as the hole gets completely localized on one side [Fig.~\ref{Fig3}(b)].

Surprisingly, we also find a narrow region (III) close to $u=0.5$ where $S_{\rm total}$ is very small. Here the hole is confined in between the two frustrated plaquettes [Fig.~\ref{Fig3}(c)] and there are two domain walls at $\pm p$ separating regions of opposite magnetization [Fig.~\ref{Fig3}(f)]. We have performed Density Matrix Renormalization Group (DMRG) simulations using the \texttt{block2} library \cite{block2} to extend the results from exact diagonalization to small values of $S_{\text{total}}$.

The ferromagnetic backgrounds are not affected by system size and we find the same phase diagram up to $l=32$. In contrast, for exchange coupling $S_{\rm total} = 0 \; \forall u$ and the plaquettes do not behave collectively (see SM \cite{SuppMat}).

{\it Grid with diagonal bonds}|To study the effect of building up frustration, we add diagonal bonds to a rectangular grid one at a time, with the same hopping $t>0$ on all bonds, and study the ground state using exact diagonalization. If the diagonals are drawn at random from a Shastry-Sutherland lattice \cite{SRIRAMSHASTRY19811069}, we find each of them reduces $S_{\rm total}$ by $1$. Even if the bonds are drawn without such a constraint, a similar behavior is reproduced [Fig.~\ref{Fig4}(a)]. Based on the collective physics of two plaquettes (region IV in Fig.~\ref{Fig3}), we expect this reduction arises from the hole being shared by all the diagonals, each of which binds a singlet. Such a structure is supported by the spin correlations [see Fig.~\ref{Fig4}(b)] and agrees with \cite{cheng_shastry_sutherland_2004} for a regular Shastry-Sutherland lattice. As before, we do not expect the physics to be limited by system size. 

It is surprising that a local change in topology would alter the collective ordering of a strongly correlated quantum system in such a predictable manner. This does not occur for exchange magnetism, where the diagonal bonds have $\langle \hat{S}_i . \hat{S}_j \rangle > 0$ (see SM \cite{SuppMat}).

{\it Random graphs}|Finally, we examine whether the role of frustration in shaping the magnetic order extends to random Erd\H{o}s--R\'enyi graphs with a given number of sites and bonds. While such graphs do not have well-defined frustration centers, their frustration level can be measured by the number of odd loops or by the frustration index~\cite{PhysRevE.68.056107, aref2018measuring, aref2019balance}, which is the minimum number of bonds one has to remove to make the graph bipartite. From exact diagonalization for $20$-site graphs (Hilbert-space size $\sim 2 \times 10^6$), we find a strong negative correlation ($\sim -0.5$) between $S_{\rm total}$ and the frustration index (Appendix C). In addition, $S_{\rm total}$ is positively correlated with the numbers of even loops and anticorrelated with the numbers of odd loops [Fig.~\ref{Fig5}(a)]. In fact, these correlations also persist at the level of individual bonds [Fig.~\ref{Fig5}(b)], showing the spin alignment across a bond is directly related to how frustrated that bond is (see Appendix C for other measures and SM \cite{SuppMat} for other system sizes). The correlations get weaker with increasing loop size as a large loop generically overlaps with many smaller loops.

With the above insight, one can tune the magnetization of random graphs by limiting the minimum loop size (``girth'') $r_{\rm min}$ \cite{bayati2018generating}. As $r_{\rm min}$ is increased, the average value of $S_{\rm total}$ as well as its distribution swings back and forth [Figs.~\ref{Fig5}(c)-\ref{Fig5}(e)], moving to higher (lower) values with the removal of odd (even) loops, until one reaches the maximum girth \cite{Verstrate2016}. If this is even, the graphs become close to bipartite and ferromagnetic, whereas if it is odd, they are strongly frustrated with a small $S_{\text{total}}$. None of these variations are present for a Heisenberg model (SM \cite{SuppMat}).

{\it Conclusions}|In summary, the interplay of strong interactions, kinetic frustration, and nonuniform topology gives rise to the physics of singlet binding, allowing one to shape the magnetic order of strongly correlated fermions by embedding frustration. This is fundamentally different from exchange magnetism and has no classical analog.

While it is challenging to realize Hubbard models with arbitrary connectivity, one can use well-established techniques to locally turn on diagonal hops for cold atoms in optical lattices. Such systems offer a controllable testbed for the Hubbard model \cite{Esslinger2010, gross2017quantum, review_coldatoms_2018} and was used recently to observe signatures of kinetic magnetism at the single-hole limit with up to $U/t = 72$ \cite{experiment_greiner_2024, experiment_bakr_2024, Ji2021}. As we quantify in Appendix D, one can mediate diagonal hops via an intermediate site at the center of the selected plaquette by applying a focused laser beam. 
The energy scales can be tuned further by modulating~\cite{nixon2024individually} or staggering \cite{Chalopin2025} the on-site potentials. More general connectivities should be possible for limited number of sites using digital quantum simulation \cite{Lloyd1996, BassmanOftelie2021} with programmable atom arrays \cite{GonzlezCuadra2023} and superconducting qubits~\cite{fermionsimul_transmons_2015, Lamata2018}.

Our findings motivate further studies of kinetic magnetism on graphs that can explore (i) deeper connections to polaron physics \cite{demler_attraction_2024, samajdar_polaronic_2024} and effective spin models \cite{haerter_shastry_triangular_2005, numerical_manuel_2014}, (ii) the physics of multiple holes, which is even richer and sensitive to the underlying geometry \cite{doucot_wen_1989, shastry_variational_1990, riera_1989, ivantsov_2017, xavier_ring_2020, demler_attraction_2024, zhang_pairing_2018, sujun_2023, white_kivelson_2012, vishwanath_triangular_2022, hitesh_square_triangular}, (iii) string orders arising from the competition of exchange and kinetic magnetism \cite{laughlin_string_1996, string_experiment_2019, demler_string_2019, demler_attraction_2024, grusdt_crossover_2024}, (iv) importance of other graph properties such as separability and heterogeneity \cite{tasaki1989extension, 15puzzle_2018, decorated_chains_1999, site_removal_2024, mielke_line_graphs_1991, preethi2024frustrated, network_entanglement_2025}, (v) effect of thermal fluctuations \cite{demler_highT_2023, khatami_finiteT_2025, mueller_2025}, and (vi) broader strong-correlation physics of the doped Hubbard model \cite{highTc_rmp_2006} on graphs.

More generally, our results show how one can harness quantum effects on networks to control collective behavior, strongly encouraging future studies at the intersection of network science and quantum condensed matter.

\begin{acknowledgments}
\vspace{1em}
{\it Acknowledgments}|We thank Alex Wietek, Roderich Moessner, Claudio Castelnovo, Ajit Balram, Kaden Hazzard, Mohit Randeria, and Subhro Bhattacharjee for stimulating discussions. This research was supported by the ANRF Prime Minister Early Career Research Grant (Project Number ANRF/ECRG/2024/005069/PMS).

\vspace{1em}
{\it Data availability}|The data and source codes supporting the findings of this study are openly available \cite{dataset, source_codes}.
\end{acknowledgments}

%


\twocolumngrid

\onecolumngrid
\vspace{1em}
\begin{center}
    \textbf{End Matter}
\end{center}
\vspace{1em}
\twocolumngrid

\appendix

\begin{figure}[h]
    \includegraphics[width=0.96\columnwidth]{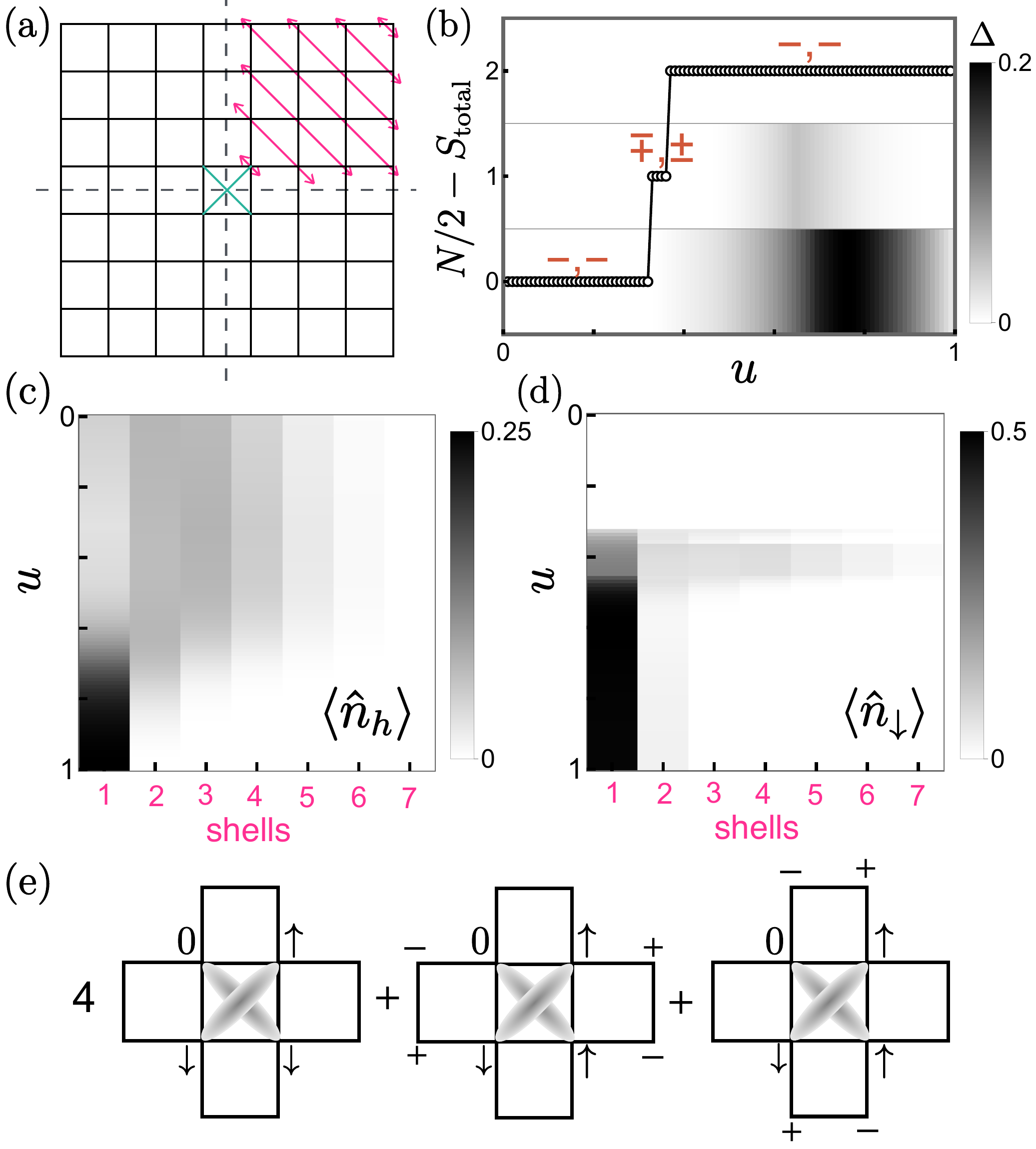}
    \caption{(a) Square grid with a central frustrated plaquette. The pink arrows define ``shells'' composed of sites at a given hop distance from the center. (b) Minimum energy sector and ground-state parity, with the gray background showing the excitation gap. (c) Hole occupation and (d) $\downarrow$-spin occupation of different shells. (e) Sketch of the ground state for large $u$. }
    \label{Fig6}
\end{figure}

{\it Appendix A: Square grid with a frustrated plaquette}|The ordering of a square grid with a central frustrated plaquette is similar to that of the ladder, as shown in Fig.~\ref{Fig6}. Compared to the ladder [Figs.~\ref{Fig2}(b)-(d)], region II with $(\pm,\mp)$ parity is smaller, the $\downarrow$-spin density is more delocalized in II and more tightly bound to the center in III. Third-order perturbation in $t/t^{\prime}$ shows that the latter is described by a $4$:$1$:$1$ superposition of three cross-dimer states, as sketched in Fig.~\ref{Fig6}(e): (i) $\ket{\times}_{\downarrow}$, (ii) $\ket{\times}_{\uparrow,(1,0)}$ where the second $\downarrow$ spin is in the neighboring horizontal plaquettes, and (iii) $\ket{\times}_{\uparrow,(0,1)}$ where it is in the neighboring vertical plaquettes. So the probability of finding a magnon outside the center is $P_{\text{out}} = \frac{1}{9}$, as opposed to $\frac{1}{5}$ for the ladder, which means they are more tightly bound even though there are more ways to delocalize. 

\begin{figure}[h]
    \includegraphics[width=0.96\columnwidth]{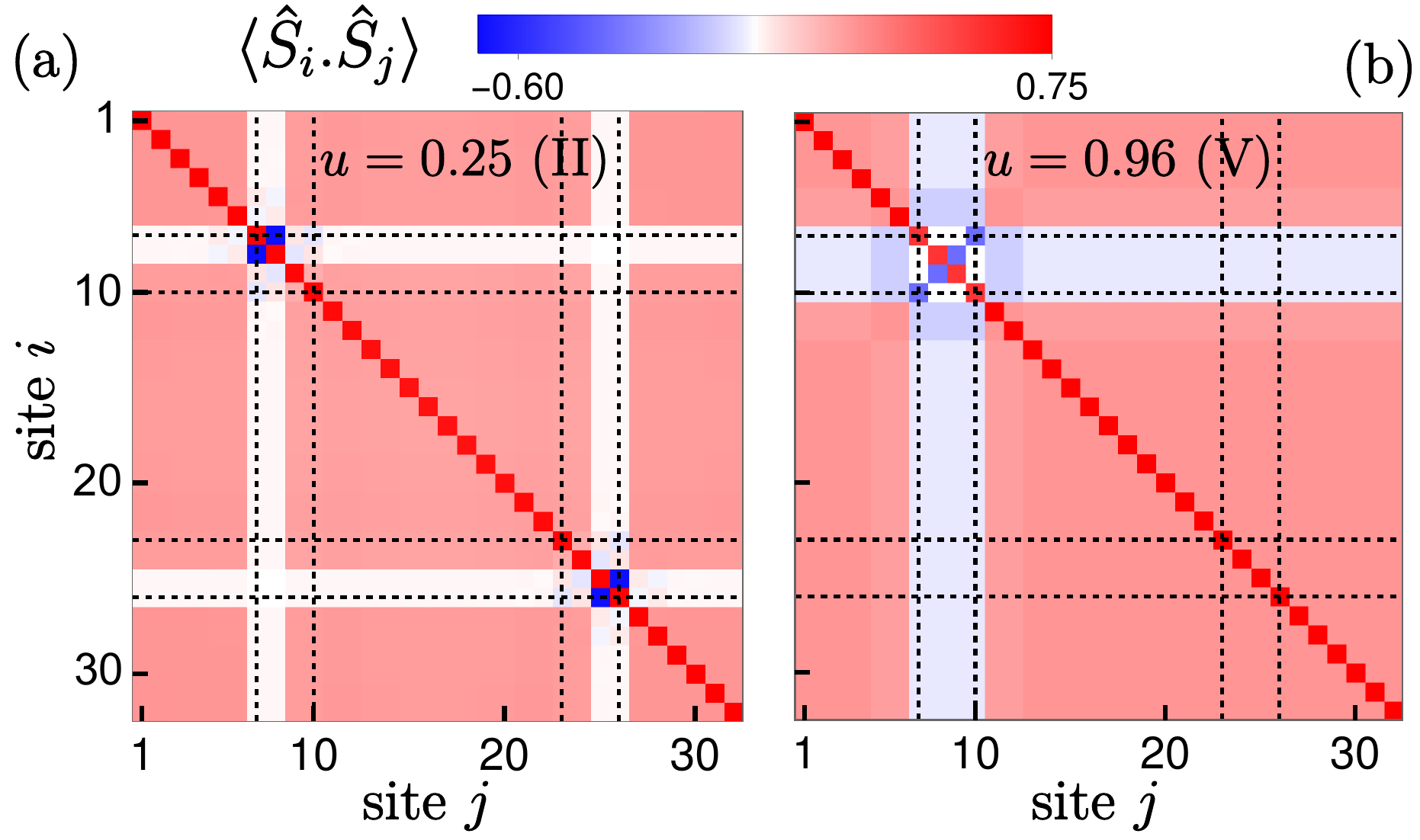}
    \caption{Spin correlations for a ladder with two frustrated plaquettes, showing vertical singlets on both in region II and diagonal singlets on one in region V [cf. Figs.~\ref{Fig3}(e)-(f)].}
    \label{Fig7}
\end{figure}

{\it Appendix B: Ladder with two frustrated plaquettes}|As explained in the main text, two frustrated plaquettes in a ladder behave independently in regions II and V of Fig.~\ref{Fig3}. Figure~\ref{Fig7} shows the spin correlations in these two regions. In II, the hole is delocalized throughout the ladder and each plaquette binds a magnon, leading to singlet correlations along the two vertical rungs that are closer to the boundary. In V, the hole is completely localized in one of the plaquettes which assumes the cross-dimer state $\ket{\times}_{\downarrow}$, while the other is spin polarized.

\begin{figure}[h]
\includegraphics[width=\columnwidth]{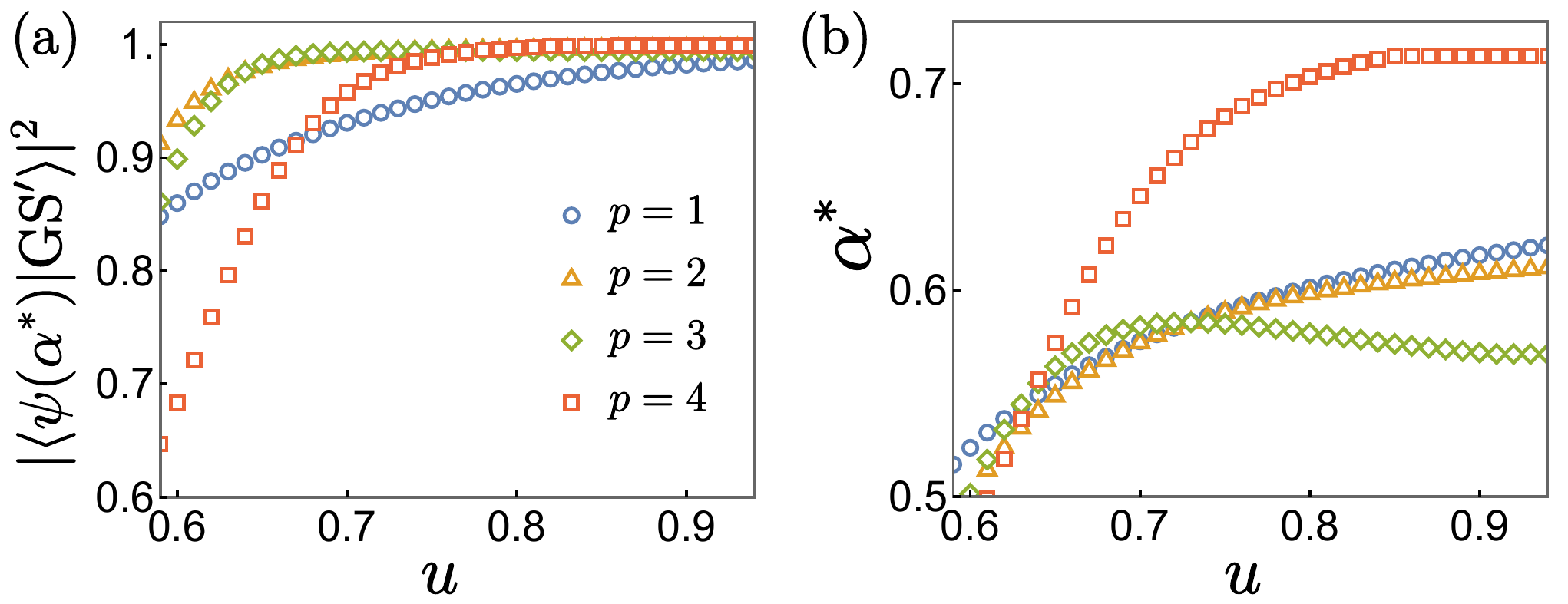}
\caption{(a) Maximum overlap $|\langle \psi(\alpha^*)|\text{GS}^{\prime}\rangle|^2$ and (b) deformation $\alpha^*$ describing singlets in region IV for $l=10$.}
\label{Fig8}
\end{figure}

In contrast, in region IV, the hole is shared by the two plaquettes, each of which binds two magnons [Fig.~\ref{Fig3}(e)]. This binding can be understood by focusing on the part of the wavefunction, $\ket{\text{GS}^\prime}$, where both the hole and the magnons are contained in the frustrated plaquettes. This is well approximated by the variational form $|\psi(\alpha)\rangle$ in Eqs.~\eqref{eq:var_wf} and \eqref{eq:distorted_singlets}. Figure~\ref{Fig8} shows that the optimal value $\alpha^*$, which maximizes the overlap $|\langle \psi(\alpha)|\text{GS}^{\prime}\rangle|^2$, is close to $0.6$ and varies slowly with $u$.

\begin{figure}[h]
    \includegraphics[width=\columnwidth]{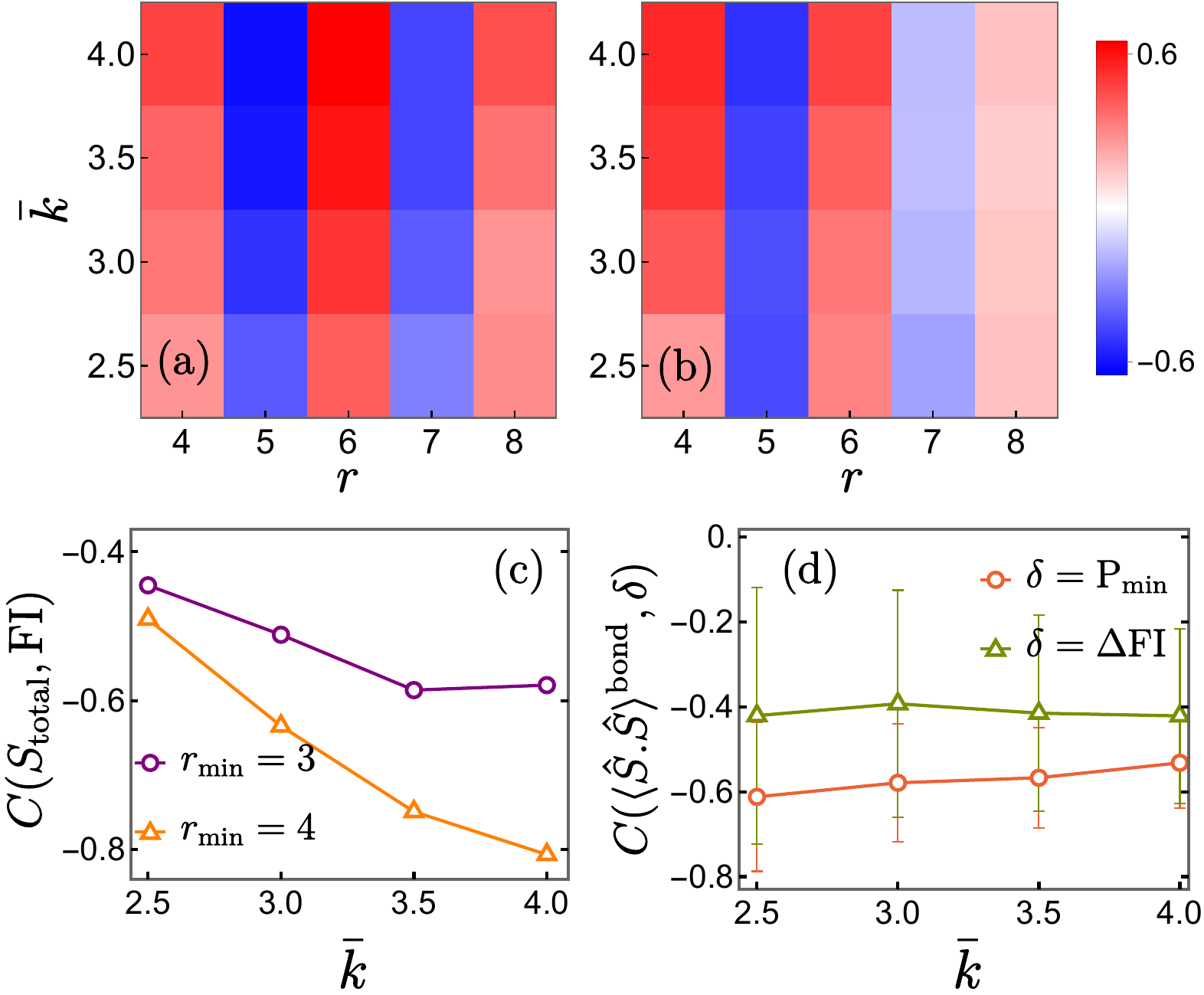}
    \caption{Correlation of (a) $S_{\rm total}$ and (b) spin alignment across individual bonds with the associated number of loops for $20$-site random graphs without triangles ($r_{\rm min}=4$) [cf.~Figs.~\ref{Fig5}(a)-(b)]. (c) Correlation of $S_{\rm total}$ with the frustration index (FI). (d) Correlation of the spin alignment across a bond with (i) parity (even = 0, odd = 1) of the smallest loop size containing the bond and (ii) change in FI due to removal of that bond, averaged over $1000$ random graphs with $r_{\text{min}} = 3$ (dots give mean values and error bars give standard deviations).}
    \label{Fig9}
\end{figure}

{\it Appendix C: Random graphs}|As shown in Fig.~\ref{Fig5} of the main text, both $S_{\rm total}$ and spin alignment across individual bonds exhibit positive and negative correlations with the numbers of even and odd loops, respectively. Figures~\ref{Fig9}(a) and (b) show that this feature is even more prominent in graphs without triangles ($r_{\rm min}=4$). In addition, Fig.~\ref{Fig9}(c) shows a strong negative correlation between $S_{\rm total}$ and the frustration index \cite{PhysRevE.68.056107, aref2018measuring, aref2019balance}, i.e., more non-bipartite graphs have lower net magnetization. No such correlation is found for exchange magnetism \cite{preethi2024frustrated}. Figure~\ref{Fig9}(d) shows a similar correlation of the spin alignment across individual bonds with (1) reduction in the frustration index due to removal of that bond and (2) odd/even parity of the smallest loop containing the bond, both of which measure how frustrated the bond is.

Note that the statistics are obtained from nonseparable graphs, for which $S_{\rm total}$ is unique in the ground state \cite{15puzzle_2018}.

\begin{figure}[h]
    \includegraphics[width=\columnwidth]{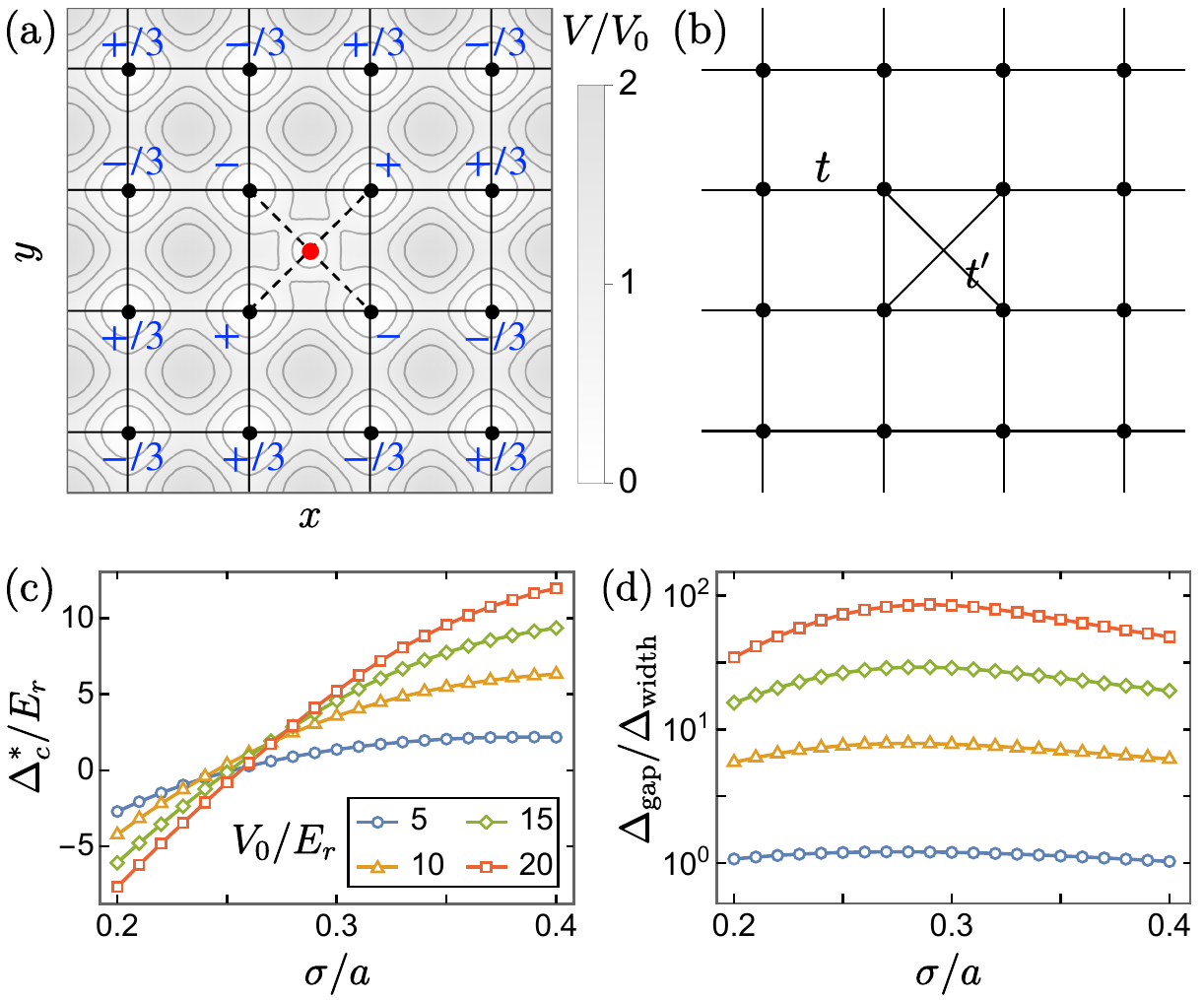}
    \caption{(a) Square lattice (black dots) with a central site (red dot), generated by the potential $V = V_0 [\cos^2(\pi x / a) + \cos^2(\pi y / a)] + (\Delta_c - 2V_0) \exp[-(x^2 + y^2) / \sigma^2]$ with $\Delta_c = 0.04 V_0$ and $\sigma = a/4$. The central site mediates diagonal hops (dashed lines). The blue labels represent amplitudes of high-frequency modulation of the on-site potentials to obtain equal hopping on all bonds. (b) Resulting graph with $t^{\prime} \approx t$. (c) Required offset $\Delta_c^*$ for equal hops in units of the recoil energy $E_r$. (d) Energy gap from the lowest band relative to the band width.}
    \label{Fig10}
\end{figure}

{\it Appendix D: Realizing diagonal hops in an optical lattice}|Here we outline a procedure to realize diagonal hopping in selected plaquettes of a square lattice for ultracold atoms. These atoms are trapped in an optical potential $V(\mathbf{r})$ created by off-resonant laser beams, which produce a dipole force proportional to the laser intensity~\cite{Grimm2000}. Using standing-wave lasers in orthogonal directions yields $V = V_0 [\cos^2(\pi x / a) + \cos^2(\pi y / a)] + V_z \cos^2(\pi z / a)$, where $a$ is the lattice spacing. One can freeze the motion along $z$ and confine the $x$-$y$ dynamics to the lowest energy band by setting $V_z \gg V_0 \gg E_r$, where $E_r \coloneqq \pi^2 \hbar^2 / (2 m a^2)$ is the recoil energy and $m$ is the atomic mass. The resulting nearest-neighbor tunneling $t_0$ decreases sharply with $V_0 / E_r$ \cite{Bloch2005}. Loading fermionic atoms with two hyperfine states realizes a Hubbard model \cite{review_coldatoms_2018} where the interaction $U$ is tunable over a wide range \cite{Bloch2008, Chin2010, experiment_greiner_2024}.

In recent years it has become possible to engineer the potential landscape within a unit cell using spatial light modulators \cite{Weitenberg2011, Zupancic2016, Lieblattice2024, Gross2021}. For diagonal hops, one can focus a gaussian beam at the center of the selected plaquette to create a local minimum with offset $\Delta_c$ [Fig.~\ref{Fig10}(a)]. It is possible to vary $\Delta_c$ and the beam width $\sigma$ such that the Wannier function at the minimum is gapped but mediates a ``virtual'' diagonal hopping $t^{\prime} \sim t_0$ [Figs.~\ref{Fig10}(c)-(d)]. 

The smaller potential barrier inside the square also increases the hopping $t_{\square}$ along its edges. To rectify this distortion, one can either apply additional repulsive potentials on the edges or, following \cite{nixon2024individually}, modulate the on-site potentials with amplitude $\pm A$ on the square and $\pm A/3$ outside, as sketched in Fig.~\ref{Fig10}(a). For high modulation frequency $\omega$, such driving rescales the non-diagonal hops to $t_{\square} J_0(\frac{2A}{\omega})$ and $t_0 J_0(\frac{2A}{3\omega})$, where $J_0$ is the Bessel function of the first kind. Thus, all bonds have equal hopping if $t^{\prime} = t_{\square} J_0(\frac{2A}{\omega}) = t_0 J_0(\frac{2A}{3\omega})$. Figures \ref{Fig10}(c)-(d) show this condition can be met while maintaining a large gap for a range of gaussian beams and $V_0 / E_r \gtrsim 5$. The hopping is close to $t_0$ and decreases slowly with $\sigma$.

Note that the procedure can be readily generalized for multiple frustrated plaquettes.


\onecolumngrid
\clearpage
\renewcommand{\baselinestretch}{1.3}\normalsize
\begin{center}
\textbf{\large Supplemental Material for\\
	    ``Shaping Magnetic Order by Local Frustration for Itinerant Fermions on a Graph''}\\
\vspace{0.5cm}
Revathy B. S${}^1$ and Shovan Dutta${}^1$

${}^1${\it Raman Research Institute, Bangalore 560080, India}
\end{center}
\vspace{0.8cm}



    \setcounter{table}{0}
    \renewcommand{\thetable}{S\arabic{table}}%
    \setcounter{figure}{0}
    \renewcommand{\thefigure}{S\arabic{figure}}%
    \renewcommand{\theHfigure}{S\arabic{figure}}
    \setcounter{secnumdepth}{3}
    \setcounter{section}{0}
    \renewcommand{\thesection}{\Roman{section}}%
    \setcounter{subsection}{0}
    \renewcommand{\thesubsection}{\Alph{subsection}}
    \setcounter{equation}{0}
    \renewcommand{\theequation}{S\arabic{equation}}%
    \setcounter{page}{1}
    \renewcommand{\thepage}{SM-\arabic{page}}
    \renewcommand{\bibnumfmt}[1]{[S#1]}
    \renewcommand{\citenumfont}[1]{S#1}


\makeatletter
\renewcommand{\baselinestretch}{1}\normalsize
\normalfont
\setstretch{1.2}

\section{Comparison with exchange magnetism}

Here we compare the results of the main text with the physics of exchange coupling on the same networks, where there is no hole and the bonds represent antiferromagnetic Heisenberg coupling $J$, as opposed to hole hopping $t$. This corresponds to the Hubbard model at half filling with large but finite on-site repulsion $U$, which gives $J \approx 4t^2/U$ \cite{supp_keeling_notes}. As we show below, in this case both the total spin and the spin alignment across individual bonds exhibit a lack of sensitivity to frustration. In particular, a frustrated bond is far from a singlet when all bonds are of equal strength. A similar insensitivity of the total spin was found for both classical and quantum spins in Ref.~\cite{supp_preethi2024frustrated}. This is in stark contrast to the doped scenario in the main paper, demonstrating that the quantum path interference of the hole and its sharing among frustration centers are key to shaping the magnetic order. Thus, the effect of frustration in the low-energy physics changes qualitatively by the introduction of the hole at large $U$. (For $U \to \infty$ the Heisenberg coupling vanishes, so there is no preferred magnetic order at half filling.)

We have used the \texttt{HPhi} library \cite{supp_HPhi1, supp_HPhi2} for exact diagonalization of the Heisenberg model.

\subsection{Ladder with one or more frustrated plaquettes}

\begin{figure}[h]
    \centering
    \includegraphics[width=\linewidth]{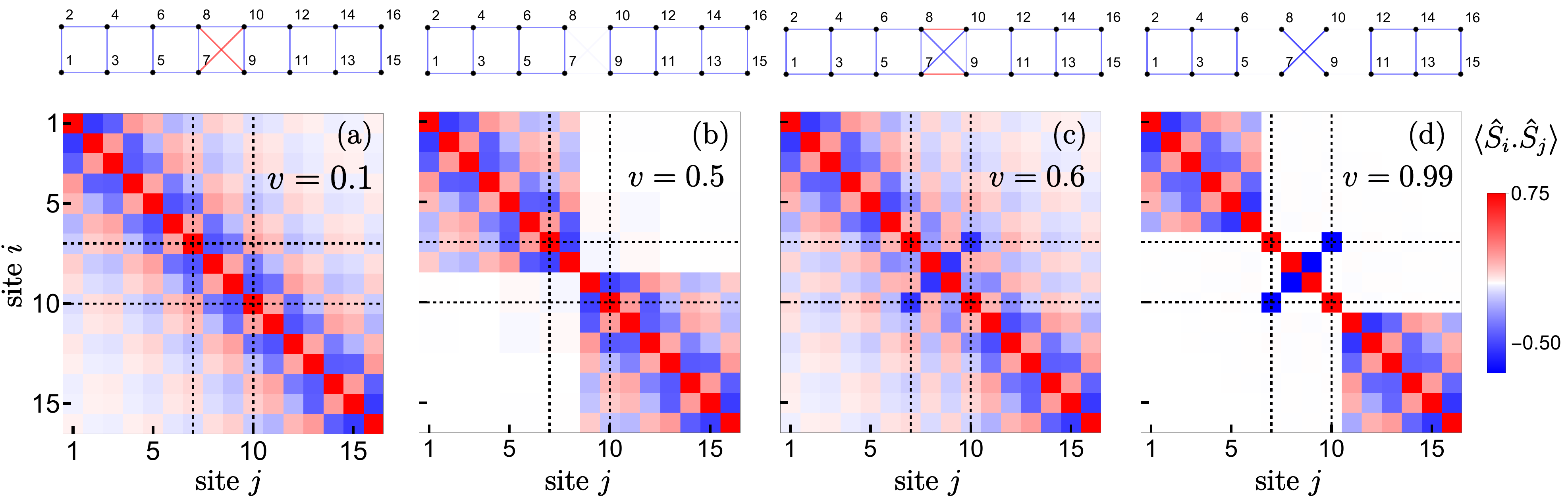}
    \caption{Ground-state spin correlations from exact diagonalization of a ladder with a frustrated plaquette having Heisenberg antiferromagnetic coupling $J^{\prime} = v$ on the diagonal bonds and $J = 1-v$ on the grid bonds. Bond correlations are visualized at the top. Dotted lines show the position of the frustrated plaquette. Similar correlations are found for longer ladders.
    }
    \label{FigS1}
\end{figure}

Figure~\ref{FigS1} shows a ladder with a fully frustrated plaquette at the center, having both diagonal bonds. In analogy to the main text (Fig.~\ref{Fig2}), we take a coupling $J^{\prime} = v$ on the diagonal bonds and $J = 1-v$ on the grid bonds. The parameter $v$ is related to the relative tunneling $u = t^{\prime} / (t + t^{\prime})$ as $v = \frac{u^2}{(1-u)^2 + u^2}$.

Firstly, $S_{\rm total} = 0$ for all $v \in [0,1)$, and the correlation across the diagonal bonds changes smoothly from positive to negative as $v$ increases beyond $0.5$. In contrast, for the doped case, $S_{\rm total}$ is maximum for $u < 0.2$ and drops by $2$ for $u > 0.5$ as each diagonal strongly binds a singlet (Fig.~\ref{Fig2}). Secondly, instead of a ferromagnetic background with localized singlets found in the doped case, here the correlations are antiferromagnetic, although their range depends on $v$ (Fig.~\ref{FigS1}). In particular, for $v=0.5$ the ladder is split into two, with almost no correlation between left and right halves. For $v > 0.5$, the halves are again correlated but with a phase slip. Finally for $v \to 1$ the diagonal bonds become singlets, leading to three uncorrelated segments. 

\begin{figure}[h]
    \centering
    \includegraphics[width=\linewidth]{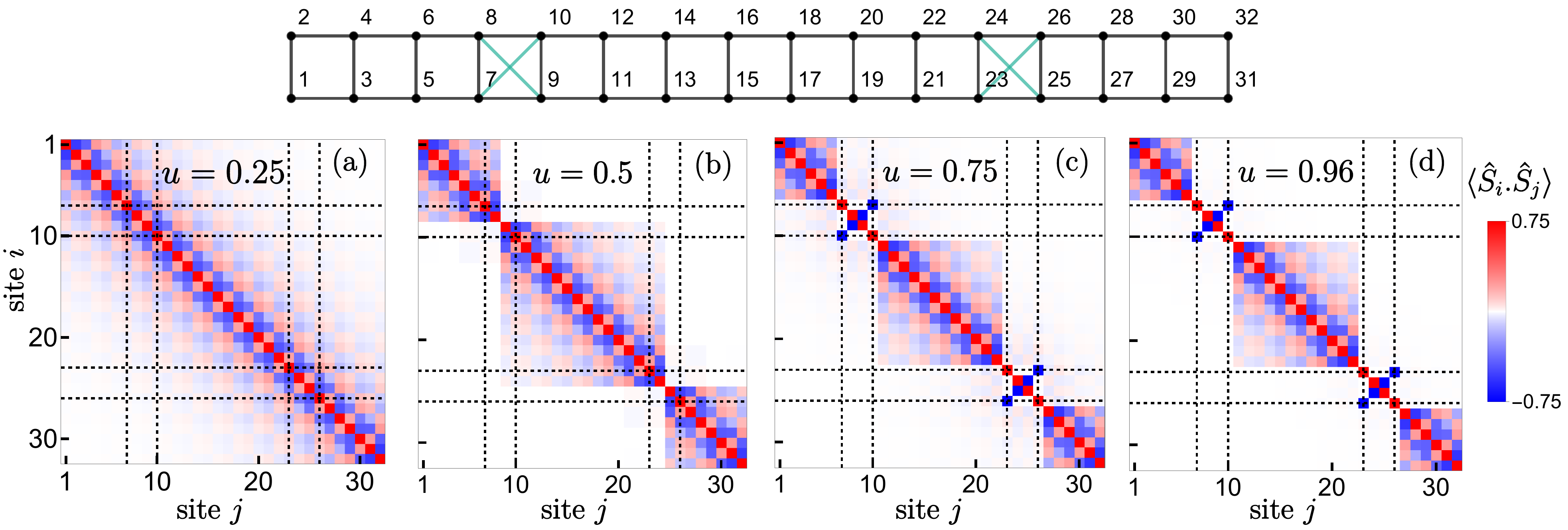}
    \caption{Correlations for a ladder with two frustrated plaquettes ($l=16, \, p=4$) with $v = u^2 / [(1-u)^2 +u^2]$ (cf.~Figs.~\ref{Fig3} and \ref{Fig7}).
    }
    \label{FigS2}
\end{figure}

For two frustrated plaquettes separated by a few sites, the physics is very similar as the plaquettes behave individually, i.e., the correlations around one are not altered by the other (Fig.~\ref{FigS2}). This leads to $3$ uncorrelated blocks for $J^{\prime} = J$ and $5$ uncorrelated blocks for $J^{\prime} \gg J$, with $S_{\rm total} = 0$ throughout. A direct comparison with Figs.~\ref{Fig3} and \ref{Fig7} shows that the physics is much richer in the doped case, where one has first-order transitions between $5$ distinct regions with very different correlations around the frustrated plaquettes. In particular, the hole can mediate strong collective effects (regions III and IV) where the ordering cannot be understood at the level of individual plaquettes.

\subsection{Grid with diagonal bonds}

\begin{figure}[h]
    \centering
    \includegraphics[width=\linewidth]{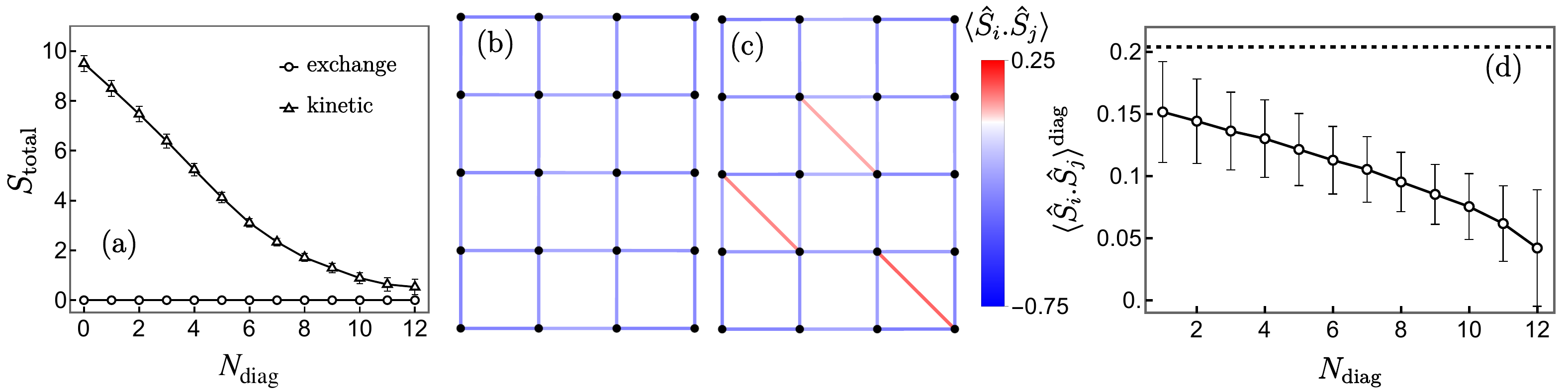}
    \caption{Total spin as a function of the number of randomly placed diagonal bonds, $N_{\rm diag}$, on a $4 \times 5$ grid (with at most one diagonal in each plaquette) for the Heisenberg model (exchange magnetism) and for the hole-doped scenario (kinetic magnetism). Dots show average values and error bars show standard deviations from up to $1000$ graphs. (b,c) Bond correlations for exchange magnetism with and without diagonal bonds [cf.~Fig.~\ref{Fig4}(b)]. (d) Average spin correlation across the diagonal bonds as a function of $N_{\rm diag}$. Error bars show standard deviations. Dotted line shows the average diagonal correlation for $N_{\rm diag} = 0$.
    }
    \label{FigS3}
\end{figure}

The contrast between exchange and kinetic magnetism is very clear for a rectangular grid with random diagonal bonds of equal strength. As shown in Fig.~\ref{FigS3}, for exchange coupling the addition of such bonds does not affect $S_{\rm total}$ (which remains zero) and only weakly reduces the spin alignment across the diagonals, which remain far from singlets. The average spin correlation across the diagonal bonds decreases slowly with the number of such bonds but remains positive [Fig.~\ref{FigS3}(d)]. On the other hand, for the doped case every diagonal bond binds a singlet, reducing the total spin by approximately $1$ [see Fig.~\ref{FigS3}(a) and Fig.~\ref{Fig4} in the main text].

\subsection{Random graphs}

\begin{figure}[h]
    \centering
    \includegraphics[width=\linewidth]{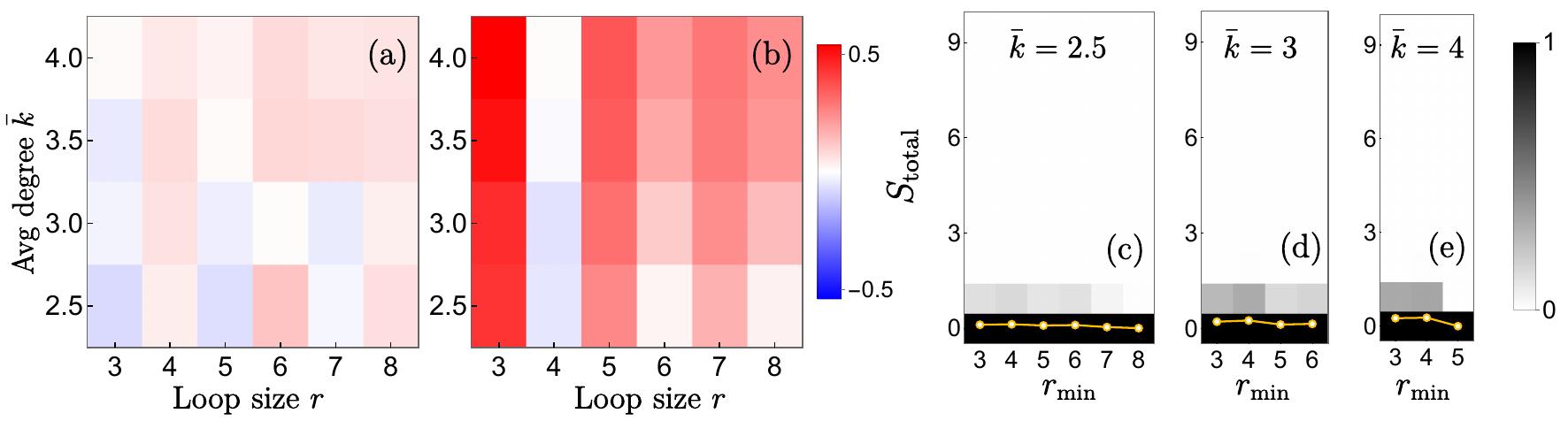}
    \caption{(a) Correlation between $S_{\rm total}$ and loop counts in $20$-site random graphs with Heisenberg antiferromagnetic bonds [cf. Fig.~\ref{Fig5}(a)]. (b) Average correlation between the spin alignment across individual bonds, $\langle \smash{ \hat{S}_i . \hat{S}_j } \rangle$, and the number of loops that the bond is a part of [cf.~Fig.~\ref{Fig5}(b)]. (c)-(e) Distribution (gray scale) and average values (yellow dots) of $S_{\rm total}$ as a function of the minimum loop size $r_{\rm min}$ [cf.~Fig.~\ref{Fig5}(c)]. Each data point is obtained from an ensemble of $1000$ nonseparable graphs.
    }
    \label{FigS4}
\end{figure}

For kinetic magnetism on random graphs we showed that both $S_{\rm total}$ and the spin alignment across individual bonds are dictated by the frustration level. In particular, both are negatively correlated with odd loops and positively correlated with even loops, which can be used to swing the total-spin distribution back and forth by increasing the minimum loop size $r_{\rm min}$ (Fig.~\ref{Fig5}). In contrast, for exchange magnetism $S_{\rm total}$ has very little correlation with the loop counts [Fig.~\ref{FigS4}(a)] (see also Ref.~\cite{supp_preethi2024frustrated}), whereas the alignment across bonds is positively correlated with odd loops and almost uncorrelated with even loops containing the bond [Fig.~\ref{FigS4}(b)]. The total-spin distribution remains peaked at zero with negligible spread for all $r_{\rm min}$ [Figs.~\ref{FigS4}(c)-(e)]. The characteristic oscillations in Figs.~\ref{Fig5}(c-e) are absent.

\section{Random graphs with different system sizes}
\begin{figure}[h]
    \centering
    \includegraphics[width=0.85\linewidth]{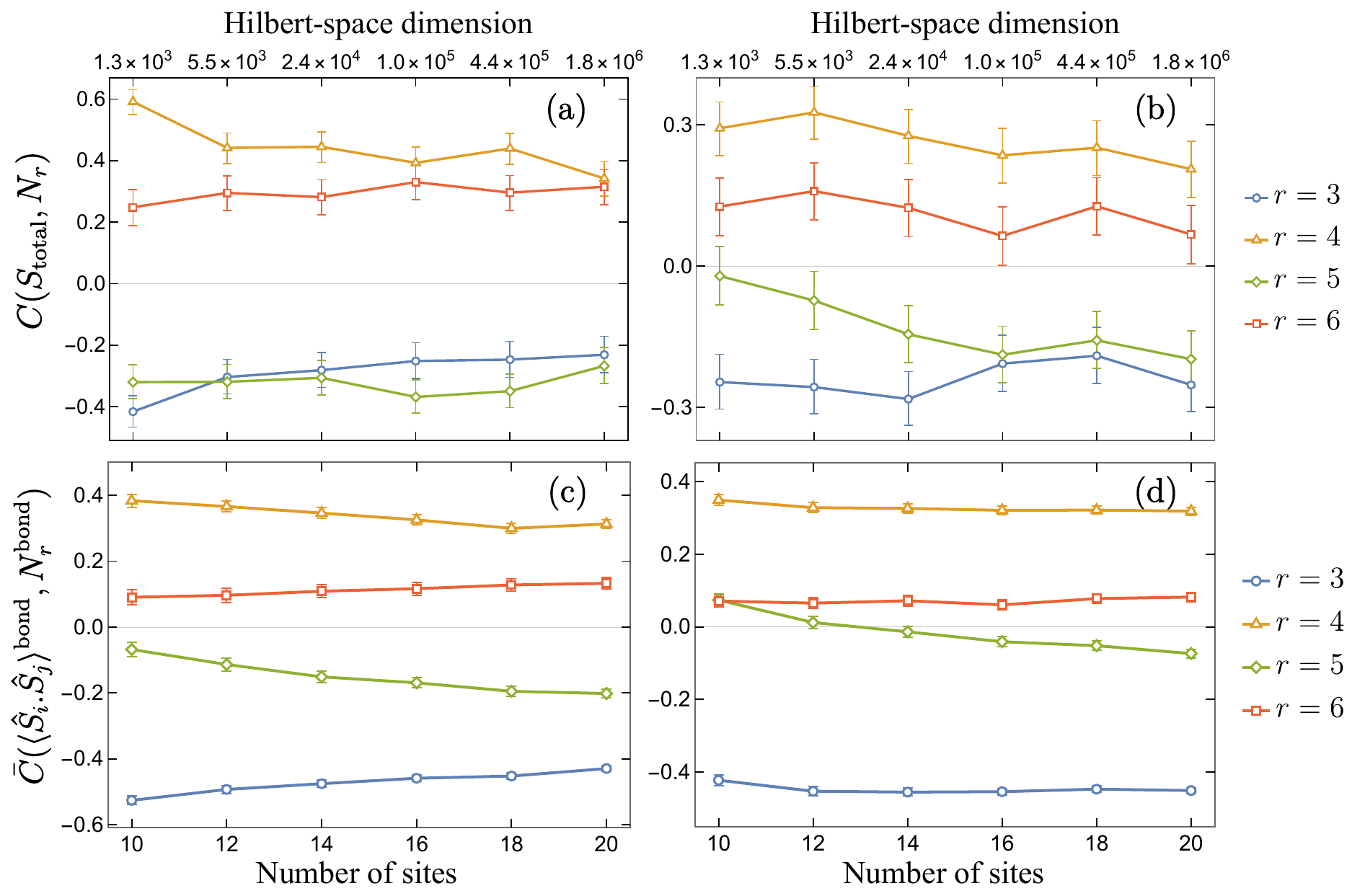}
    \caption{Top panel: Correlation between $S_{\rm total}$ and number of loops of a given size $r$ for random nonseparable graphs with varying number of sites and average degrees (a) $\bar{k} = 3$ and (b) $\bar{k}=4$. Bottom panel: Average correlation between spin alignment across individual bonds and the number of loops of size $r$ that the bond is part of, with (c) $\bar{k} = 3$ and (d) $\bar{k}=4$. Odd loops show negative correlations and even loops show positive correlations. Each dot corresponds to average over an ensemble of up to $1000$ graphs and the error bars represent $95$\% confidence intervals.
    }
    \label{FigS5}
\end{figure}

In the main text we showed that, for kinetic magnetism on random graphs, both the total spin and the bond-level spin alignment is strongly correlated with the number of loops of a given size $r$. Figure~\ref{FigS5} shows that these correlations persist for a range of system sizes.

\begin{thebibliography}{90}
\bibitem [{Sup()}]{SuppMat}%
  \BibitemOpen
  \href@noop {} {\bibinfo {title} {See {S}upplemental {M}aterial for comparisons with {H}eisenberg antiferromagnetism and data for random graphs of varying system size, which includes {R}efs.~\cite{HPhi1, HPhi2}.}}\BibitemShut {Stop}%
\end{thebibliography}
\end{document}